# Piecewise model with two overlapped stages for the structure formation and hardening upon severe plastic deformation

E. F. Talantsev[*], M. V. Degtyarev, T. I. Chashchukhina, L. M. Voronova, and V. P. Pilyugin

M. N. Mikheev Institute of Metal Physics, Ural Branch, Russian Academy of Sciences, 18, S. Kovalevskoy St., Ekaterinburg, 620108, Russia

*Abstract*

The evolution of metals micro/nano-structure upon severe plastic deformation (SPD) is still far to be theoretically explained, while experimental datasets are persistently growing. Major problem associated with understanding of SPD is a synergetic effect of several competing processes which alter material structure. In this paper we propose a model to reveal, as free-fitting parameters, strain breakpoints where predominantly one mechanism determines the micro/nano-structure in SPD materials from material hardness vs true strain experimental data. The model is applied to analyse SPD data for pure polycrystalline iron and two distinctive strain breakpoints have been revealed.

**Piecewise model with two overlapping stages for structure formation and hardening upon severe plastic deformation**

The term of severe plastic deformation (SPD) is defined a solid state material treatment in which a very large strain is applied on a bulk sample in order to make an ultra-fine grained material [1]. It should be noted, that from its beginning, SPD is intended to be a part of industrial technology [2] which covers a variety of materials ranging from pure metals [1] to ceramics [3]. Historical review and current development of SPD technology can be found elsewhere [4-7].

One of main problem associated with the development of quantitative theory of SPD is related to a fact that this materials treatment technique is a synergetic product of several competing physical processes which alter the material micro/nano-structure. In the result, materials structure is not either crystals with long range order, nor amorphous solids with short range order only, but something is in between, as this was first pointed out by Birringer *et al* [8]. Despite a fact that these authors [8] named nanocrystalline materials as "gas-like solids" (which of course has deep physical meaning), perhaps these days more accurately SPD materials can be named as frozen quantum liquid, because atoms/ions in the SPD materials are in a very strong interaction with each other (and strong particle interaction is not a characteristic of gaseous media).

Remarkably, that the first study of physical properties of SPD materials, which was well ahead of studies of mechanical properties of these materials, was a study of SPD processed quantum materials, i.e. superconducting niobium alloys reported by Fietz and Webb [9]. These authors introduced the term of "severe plastic deformation" in the literature.

One of milestone result in understanding of SPD is experimental reports on the violation of Hall-Petch relationship in ultrafine grained materials [10-13]. This experimental result is a good accord with our view that attempts to extrapolate physical laws established for microcrystalline materials on SPD materials (with minor modifications) should be failed. However, because the theory of quantum materials is far away to be developed, our current approach is to try to implement some empirical laws which have been found in the field of quantum materials for last 50 years to SPD counterparts.

From a variety of materials on which our attempt can be started, we make a choice for SPD iron. Obviously, iron is considered to be one of model metal [8] in which properties of ultra-fine grained materials can be prominently observed. From other hand, iron is a basic construction material of modern world and, thus, any improvement in iron properties will have a massive impact on global economy.

By considering a property which can be analysed in SPD iron, we take in account that the hardness is one of most conveniently measured property in SPD materials [14], and thus the establishing of quantitative relation for this property can be interesting for whole R&D field [14,15].

Thus, the purpose of this paper is to reveal a quantitative relation between hardness vs true strain in SPD iron. One of our findings is reported herein.

In this paper we analyse experimental data on SPD iron samples which have been deformed by high pressure torsion (HPT) technique. Experimental details can be found elsewhere [16].

In our previous report [16,17] we showed that some SPD metals exhibit two distinctive microstructures and hardness vs strain dependences, which are attributed to relatively low and high true strain. In Fig. 1 we present hardness vs true strain of pure iron which is taken from our previous papers [16,17] and TEM images of material structure at minimal and maximal true strain which is achievable in our experiment. Each experimental data point in Fig. 1 is the mean value of several measurements obtained on different samples subjected to as the same as different the angle of the anvil rotation. All measurements have been performed with the step of Δε = 0.2 (details can be found elsewhere [16,17]). It should be noted, that at small values of strain the iron microstructure is mainly represented by dislocation cells, which all disoriented by small angles within original grains. This material structure is significantly different from the structure after SPD treatment, for which we observed [16,17] a new ensemble of strongly disoriented microcrystals, where the boundaries of initial grains are completely reconstructed. It can be seen (Fig. 1) that despite distinctive differences in microstructures there is a difficulty to reveal characteristic dependences of hardness vs true strain for these two structural states of iron.

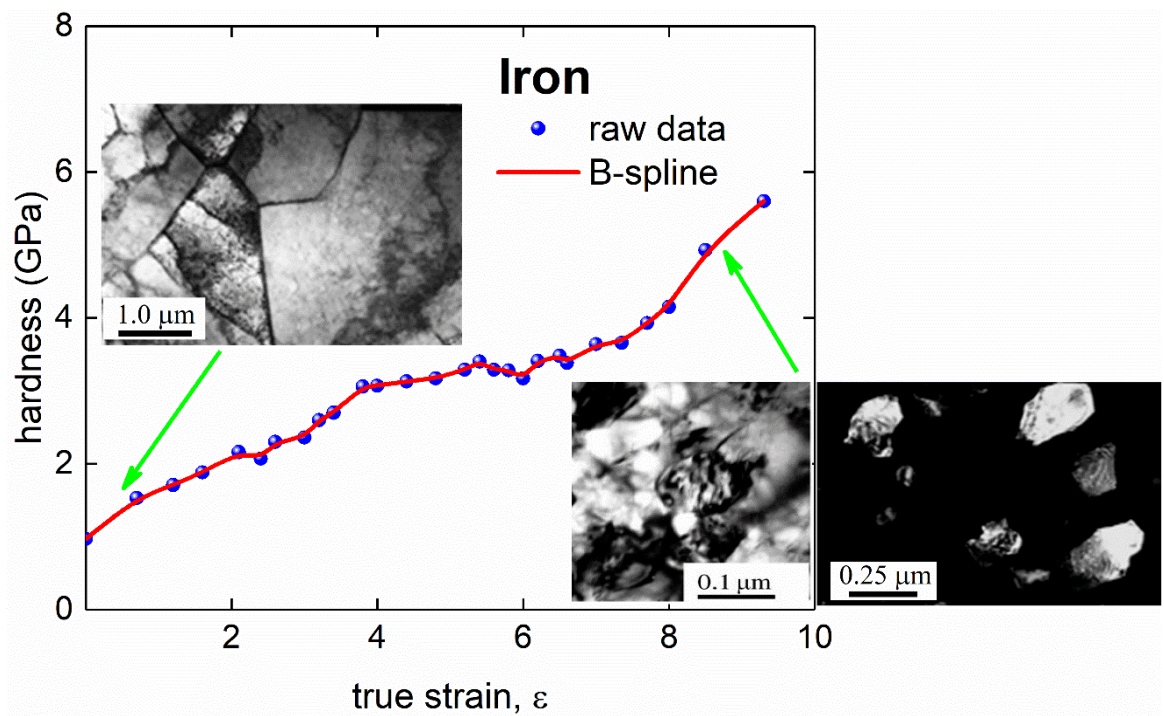

**Figure 1**. Experimental data of hardness, $H(\varepsilon)$, vs true strain and TEM images for structural stage I and structural stage II for pure iron. Raw $H(\varepsilon)$ data are taken from Refs. [16,17].

However, as this was first showed by Thompson [11] in pure nickel, that there are two distinctive grains size ranges for each of those a relation between yield strength vs grain size can be established. Later studies performed for bcc metals (iron, niobium, molibdenum) which exhibit stacking faults energy more than 140 mJ/mol showed that there are two types of structure are formed in result of HPT treatment. The structure of the first type consists of dislocation cells, and the second type is formed by microcrystallites [14,16,18-21]. It is well known, that the formation of dislocation cells is due to the motion and interaction of individual dislocations, while microcrystallites are formed due to rotational deformation modes with the participation of collective disclination effects. Based on these findings, in bcc metals deformed at room temperature, two physical mechanisms of structure formation are realized.

Thus, our model postulates that:

1. there are two distinctive deformation mechanisms which determine material hardness;
2. there is a true strain range where two mechanisms coexist;
3. in this coexisting range of true strain, the total hardness is linear additive sum of each mechanism.

These three postulates are expressed in general continuous piecewise fitting function we propose in this paper:

$$H(\varepsilon) = \theta(\varepsilon_2 - \varepsilon) \cdot H_1(\varepsilon) + \theta(\varepsilon - \varepsilon_1) \cdot H_2(\varepsilon) + \theta(\varepsilon - \varepsilon_2) \cdot \theta(\varepsilon_1 - \varepsilon) \cdot \left( \frac{|\varepsilon - \varepsilon_1|}{|\varepsilon_2 - \varepsilon_1|} \cdot H_1(\varepsilon) + \frac{|\varepsilon - \varepsilon_2|}{|\varepsilon_2 - \varepsilon_1|} \cdot H_2(\varepsilon) \right), \quad (1)$$

where $H_1(\varepsilon)$ and $H_2(\varepsilon)$ are hardness functions for structural states I and II respectively (i.e. within strain range of $(0, \varepsilon_1)$ and $(\varepsilon_2, \infty)$ respectively), $\varepsilon_1$ is free-fitting upper strain limit for the exhibiting structural state I, and $\varepsilon_2$ is free-fitting nucleation breakpoint of structural state II.

For clarity, in Fig. 2 we show the weight function of:

$$y = \theta(\varepsilon_2 - \varepsilon) + \theta(\varepsilon - \varepsilon_1) + \theta(\varepsilon - \varepsilon_2) \cdot \theta(\varepsilon_1 - \varepsilon) \cdot \left( \frac{|\varepsilon - \varepsilon_1|}{|\varepsilon_2 - \varepsilon_1|} + \frac{|\varepsilon - \varepsilon_2|}{|\varepsilon_2 - \varepsilon_1|} \right) \equiv 1 \quad (2)$$

which is splatted in two parts of structural state I and structural state II.

As we already mentioned above, our approach is based on an attempt to implement some empirical laws which have been established for quantum materials (and other complicated physical phenomena, where continuous piecewise function approach is a feasible way to describe the system [22]), for the SPD materials. One of the most widely used approach in quantum materials (and in superconductivity) is to find a scaling law of key property vs one major parameter. In mentioned above paper by Fietz and Webb [9], authors proposed to use a scaling in form of power law for the pinning force $F_p$ vs reduced magnetic field $B/B_{c2}$. Further developments of this approach have been reported since then [23-29], as well as the decomposition of complicated temperature dependences of physical properties in superconductors and normal conductors in a sum of reduced polynomic values [23,24,30-32].

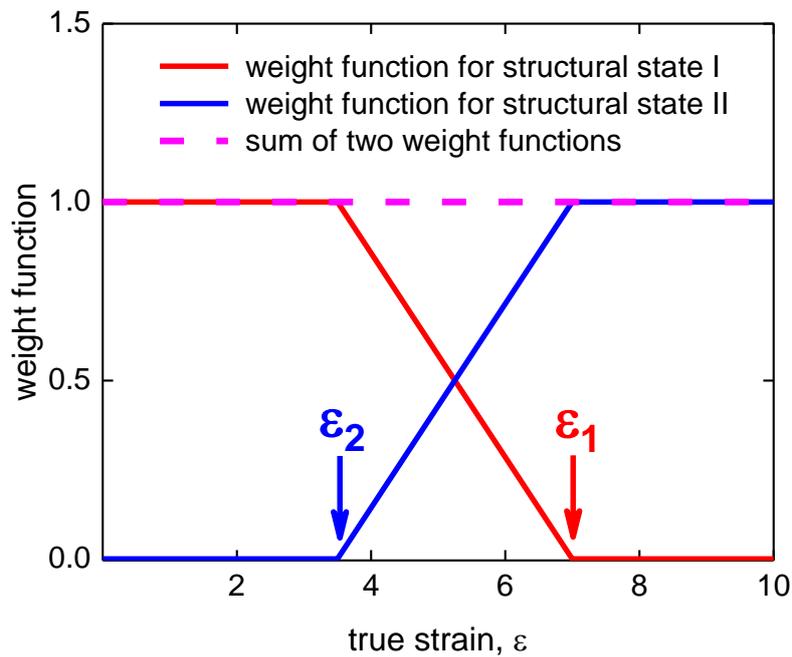

**Figure 2**. Schematic representation of weight function (Eq. 2) of the model (Eq. 1). $\varepsilon_1$ and $\varepsilon_2$ are free-fitting breakpoints of the model.

Thus, it is very convenient to use analytical form for hardness vs strain function in a form:

$$H_1(\varepsilon) = H_1 \cdot \left(1 + \left|\frac{\varepsilon}{P}\right|^{\alpha}\right), \; stage \; I$$
$$H_2(\varepsilon) = H_2 \cdot \left(1 + \left|\frac{\varepsilon}{Q}\right|^{\beta}\right), \; stage \; II$$
(3)

where $H_1$, $H_2$, $P$, $Q$, $\alpha$ and $\beta$ are free fitting parameters of the model (Eq. 1).

Experimental $H(\varepsilon)$ data and fit to Eqs. 1,3 are shown in Fig. 3. Deduced values for $\varepsilon_1$ breakpoint (at which the cellular structure, designated by stage I, disappears) and the nucleation breakpoint for stage II, $\varepsilon_2$, are:

$$\varepsilon_1 = 5.4 \pm 1.0 \tag{4}$$

$$\varepsilon_2 = 3.0 \pm 0.1 \tag{5}$$

One of the most interesting result revealed by the fit is that extrapolated $H_2(\varepsilon)$ curve (Eq. 3), which is shown in Fig. 3,b (by letter B), which has starting (nucleation) breakpoint at $\varepsilon_2 = 3.0 \pm 0.1$ has absolute value for hardness, $H_2(\varepsilon = 3.0 \pm 0.1) = 4.4 \pm 0.1\ GPa$, which is more or less equal to ultimate strength of strongest steel fibre wire made by drawing (tradename of Scifer) [33]. This means that, newly nucleated microcrystals in SPD iron at stage-II exhibits maximal strength properties.

It should be also noted, that deduced power law exponents, α and β:

$$\alpha = 0.63 \pm 0.13 \cong \frac{2}{3} \tag{6}$$

$$\beta = 2.1 \pm 0.3 \cong 2 \tag{7}$$

are in a good agreement with theoretical predicted values for SPD body-centred cubic metals [34].

Our previous approach [16] to study structural stages in HPT processed iron is based on transmission electron microscopy and the analysis of hardness vs true strain dependence for which we employed a technique proposed in Ref. [34]. The technique is to plot hardness data vs square root of true strain on which two or three linear parts (depends on material) can be distinguished. We show data of Figs. 1,3 represented in this way in Fig. 3(c) in Ref. [16]. It can be seen in Fig. 3(c) in Ref. 15, that SPD iron has three stages. The linear part at low strain region ($\varepsilon \leq 4.0$) represents the first stage of the HPT treatment at which the cellular structure is formed. Strain boundaries for this stage was determined as $\varepsilon \leq 4.0$ [16]. The highest true strain value for this region, ε = 4.0, has been designated as e$_{1-2}$ in Ref. [16].

The second linear part (Fig. 3(c) in Ref. [16]) was called the mixed-type structure and one covers the range of true strain of $4.0 \leq \varepsilon \leq 6.0$ [16]. At this deformation, the iron microstructure is represented by the dislocation cells and microcrystallites. At the largest true strain, with $\varepsilon \geq 6.0$, the iron microstructure is formed by microcrystals only. The breakpoint between the second and the third linear parts, i.e. $\varepsilon = 6.0$, was designated by e$_{2\text{-}3}$ [16].

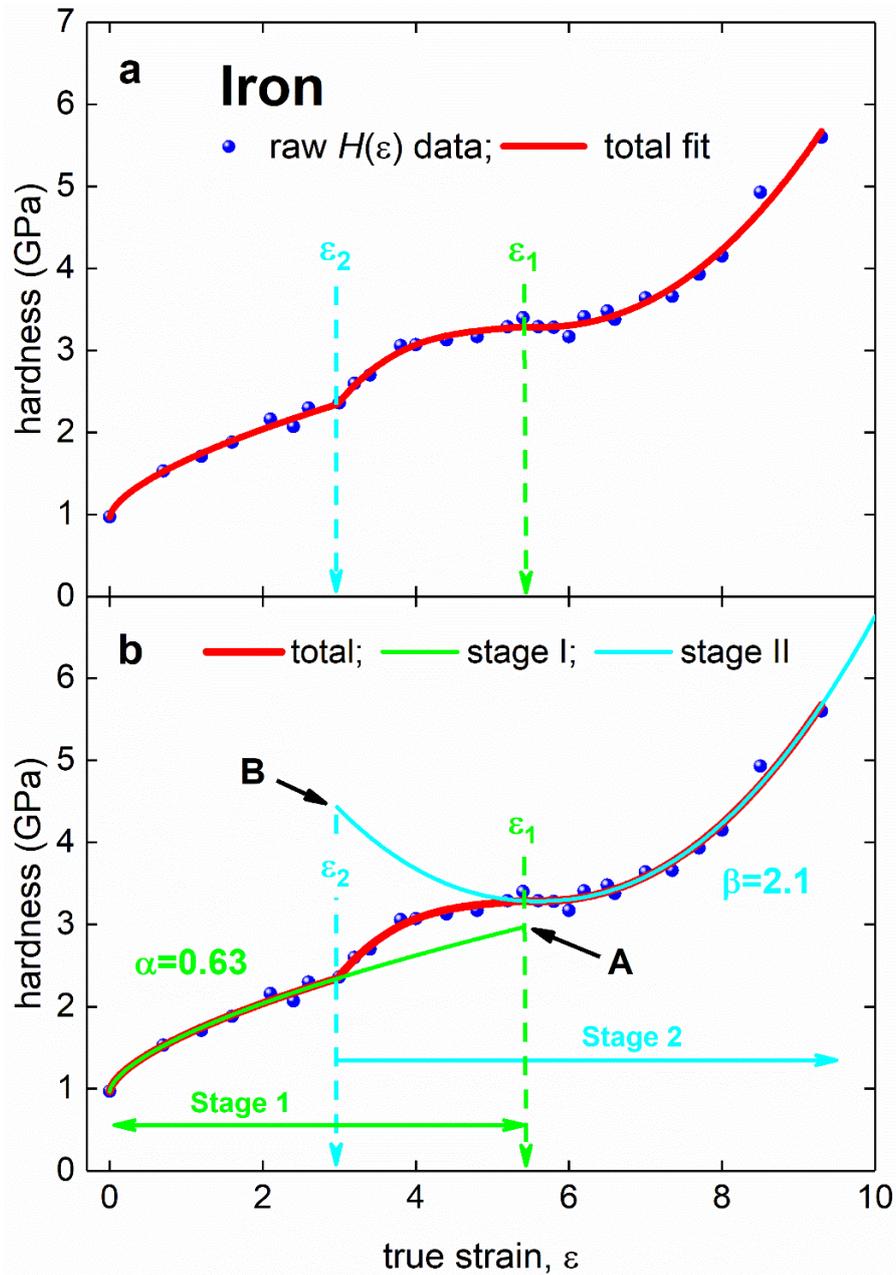

**Figure 3**. Experimental data of hardness, $H(\varepsilon)$, vs true strain and fit to Eqs. 1,3 for pure iron. (a) Total fit, breakpoints $\varepsilon_1$ and $\varepsilon_2$, and 95% confidence bars are shown; the fit quality is $R = 0.994$. (b) Fit where contributions from both stages in overlapped strain range and power law exponent are shown.

It is interesting to compare the breakpoints deduced in our previous papers [16,17] with ones deduced in this work. For instance, there is a reasonable agreement between $e_{2-3} = 6.0 \pm$ 0.2 [16] and respectful value of $\varepsilon_1 = 5.4 \pm 1.0$ deduced in this paper. It should be mentioned that this breakpoint corresponds to the HPT treatment at which dislocation cells are disappeared and iron structure is transformed in pure SMC structure [16,17].

Due to there is a much less agreement was found for the breakpoints of $e_{1-2} = 4.0 \pm 0.2$ (it is shown in Fig. 3(c) of Ref. [16]) and $\varepsilon_2 = 3.0 \pm 0.1$ (deduced in this work) we performed detailed TEM studies of iron deformed with the true strain of $\varepsilon = 3.2 \pm 0.2$ to confirm/disprove the position of $\varepsilon_2 = 3.0 \pm 0.1$ for the breakpoint. In result, we find that iron structure deformed at $\varepsilon = 3.2$ (Fig. 4) have small, but detectable amount of predicted microcrystallites. It is obvious that our primary experimental technique (i.e., hardness vs true strain), due to its integrated nature cannot directly detect a small amount of new elements of structure, i.e. microcrystallites. In addition, a disagreement between $\varepsilon_2$ and $e_{1-2}$ breakpoints is also related to facts that the model utilized in Ref. [16] uses fixed power law exponent of 2 (and one is not free-fitting parameter) and it does not use the stages overlapping idea, which both implemented in the proposed model. Thus, we find that our model is a new instructive research tool to determine strain breakpoints with much better accuracy.

In overall, in this paper we propose a continuous piecewise model with two overlapped stages for the analysis of experimental hardness vs true strain data for SPD processed materials. The novelty of this approach is lied in its ability to deduce, as free-fitting parameters, strain breakpoints which separate different micro/nano-structure modes generated upon SPD process. We have applied the model to analyse experimental data for polycrystalline samples of pure iron. The analysis reveals distinctive strain boundary of $\varepsilon = 3.0 \pm 0.1$ at which new submicrocrystalline structure starts to form in iron. Precise TEM research confirms the correctness of proposed model.

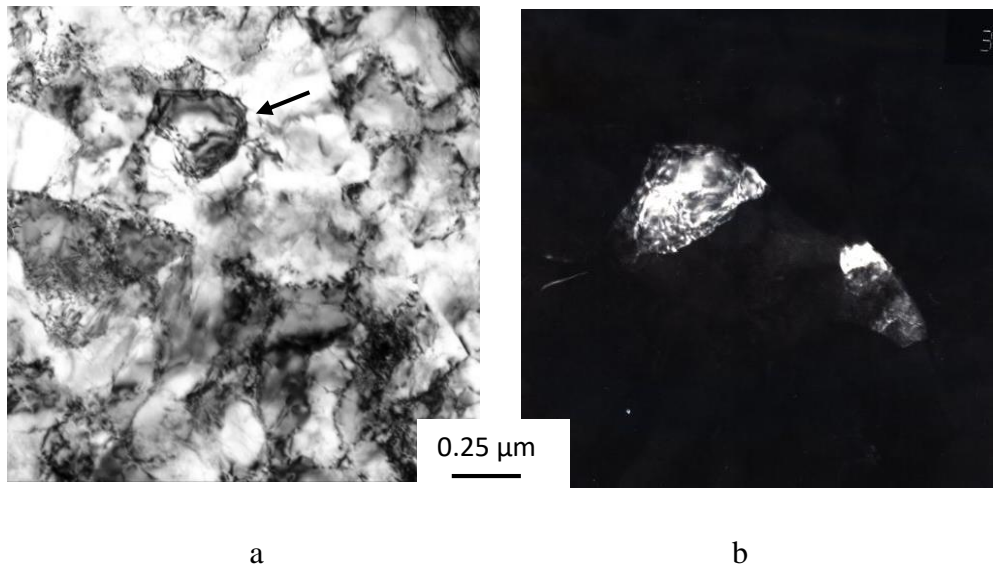

| a | b |

**Figure 4.** Iron microstructure after the deformation with ε = 3.2±0.2: a – bright field image, b – dark field image taken in the reflection (110)$_α$. The element of material structure which is designated in main text as the microcrystallite is marked with an arrow in the panel a.

**Data availability**

The datasets measured and generated during and/or analysed during the current study are available from the corresponding author on reasonable request.

**Acknowledgement**

Authors thank financial support provided by the state assignment of Minobrnauki of Russia (theme "Pressure" No. AAAA-A18-118020190104-3).